# Evolution of Quantum Computing: A Systematic Survey on the Use of Quantum Computing Tools

Paramita Basak Upama*, Md Jobair Hossain Faruk†, Mohammad Nazim‡, Mohammad Masum§, Hossain Shahriar§
Gias Uddin¶, Shabir Barzanjeh‖‡, Sheikh Iqbal Ahamed∥, Akond Rahman‡

*Department of Computer Science and Engineering, Eastern University, Bangladesh
†Department of Software Engineering and Game Development, Kennesaw State University, USA
‡Department Computer Science; Kennesaw State University, USA
§Department of Information Technology, Kennesaw State University, USA
¶Schulich School of Engineering, University of Calgary, Canada
‖‡Department of Physics and Astronomy, University of Calgary, Canada
∥Department of Mathematics, Statistics and Computer, Marquette University, USA
‡Department of Computer Science, Tennessee Tech university, USA

{*paramita.et}@easternuni.edu.bd, {†mhossa21, ‡mnazim}@students.kennesaw.edu, {§hshahria}@kennesaw.edu
{¶gias.uddin, ‖‡shabir.barzanjeh}@ucalgary.ca, {∥sheikh.ahamed}@mu.edu, & {‡arahman}@tntech.edu

*Abstract*—Quantum Computing (QC) refers to an emerging paradigm that inherits and builds with the concepts and phenomena of Quantum Mechanic (QM) with the significant potential to unlock a remarkable opportunity to solve complex and computationally intractable problems that scientists could not tackle previously. In recent years, tremendous efforts and progress in QC mark a significant milestone in solving real-world problems much more efficiently than classical computing technology. While considerable progress is being made to move quantum computing in recent years, significant research efforts need to be devoted to move this domain from an idea to a working paradigm. In this paper, we conduct a systematic survey and categorize papers, tools, frameworks, platforms that facilitate quantum computing and analyze them from an application and Quantum Computing perspective. We present quantum Computing Layers, Characteristics of Quantum Computer platforms, Circuit Simulator, Open-source Tools Cirq, TensorFlow Quantum, ProjectQ that allow implementing quantum programs in Python using a powerful and intuitive syntax. Following that, we discuss the current essence, identify open challenges and provide future research direction. We conclude that scores of frameworks, tools and platforms are emerged in the past few years, improvement of currently available facilities would exploit the research activities in the quantum research community.

## I. INTRODUCTION

Quantum computing relies on properties of quantum mechanics to compute problems that are beyond the reach of the existing classical computers and achieved significant advancements in the past few years [1], [2]. Intersecting of various techniques including physics, mathematics, computer science, and information theory paved to initiate a perfect domain, quantum computing capable of performing calculations deemed unachievable for classical computers. Compared to classical computers, quantum computers have high computational power, less energy consumption, and exponential speed [3]. It is attainable by controlling the behavior of tiny physical objects or microscopic particles including atoms, electrons, and photons that transfer digital information.

In quantum computing, zero or one bit (one-bit) of information is encoded using two orthogonal states of a microscopic object known as quantum bit or qubit [4]. Having both 0 and 1 simultaneously as its value is called "superposition". Also, they have a property known as "entanglement" and based on the changing the state of one Qubit also changes the state of another, even residing at a distance. Qubits acquire both digital and analog nature that brings the quantum computers into tremendous computational power [5]. Several quantum algorithms have already been developed, Grover's algorithm for searching and Shor's algorithm for factoring large numbers in popular [6].

TABLE 1: DIFFERENCES BETWEEN QUANTUM COMPUTING AND CLASSICAL COMPUTING

| Quantum Computing | Classical Computing |
|---|---|
| Calculates with Qubits, that can have values 0 o1 or both simultaneously | Calculates with transistors, that can have values either 0 or 1 |
| Power increases exponentially in proportion to the number of Qubits | Power increases linearly with the number of transistors |
| Have high error rates | Have lower error rates |
| Operates at close to absolute zero temperature | Operates at room temperature |
| Much secured to work with | Less secured to work with |
| Suited for big/complex tasks, such as-optimization problems, data analysis and simulations | Suited for everyday processing tasks |

Quantum programming languages are essential to translate complex ideas into instructions to be executed by a quantum computer. They facilitate the discovery and development of new quantum algorithms, as well as executing the existing ones [7]. There is a number of key differences between Quantum Computing and Classical Computing and provided in Table 1.

Quantum algorithms are already being applied in a variety of industries including healthcare, finance, manufacturing, cybersecurity, and blockchain. Optimization problems for scheduling and route planning, search algorithms, sampling and pattern matching, quantum encryption are a few of them. In healthcare, accelerating drug discovery, drug design, optimizing therapy/treatment, probable time to market new drugs are possible due to Quantum Computing in healthcare industries. Drafting trading strategies and detecting market instability for financial services seems to be plausible because of it too. Besides, advertisements strategy and product marketing, software verification, and validation are much easier with emerging Quantum Computing.

The main motivations for this study are the supreme nature of quantum computing, its advantages in solving real-world problems much more efficiently than classical computing and identifying open challenges of this emerging research field. The contribution of this paper is two-folds:

- We conduct a systematic review and present the progress of quantum computing.
- We identify the related frameworks, tools, and platforms that facilitate quantum computing.

The rest of the paper is organized as follows: In Section II, we discuss related work for Quantum Computing evolution and Tools followed by research methodology in Section III. Section IV provides details of various platforms and tools for Quantum Computing followed by explaining circuit simulators of quantum computing in Section V. Section VI explains tools and software while Section VII discusses the current situation of the field and its challenges followed by providing recommendations for future research in Section VIII. Finally, Section IX concludes the paper.

## II. RELATED WORK

Quantum computing is trying to optimize algorithms in various fields of computers by implementing and harnessing the power of qubits in a quantum environment or computer. Quantum evolutionary algorithm (QEA) and the divide-and-conquer idea of cooperative coevolution evolutionary algorithm (CCEA) are used to overcome the low solution efficiency, insufficient diversity in the later search stage, slow convergence speed and a higher search stagnation possibility of differential evolution (DE).

Six algorithms in solving six test functions from CEC'08 under the dimensions of 100, 500 and 1000 resulted in an improved differential evolution (HMCFQDE) with higher convergence accuracy and stronger stability [8]. Quantum evolutionary concepts were executed by quantum superposition and entanglement which showed a logarithmic growth rate of the number of evaluations of fitness functions needed to identify a sufficiently accurate solution, and the depth of its quantum circuits is O (1) with a significant impact on the effect of quantum noise on computation.

Google's classical PageRank algorithm has also explored its quantum implementations with the growing content being uploaded online [9]. The Quantum PageRank algorithm was simulated on a six node web network for observation against the PageRank algorithm. Quantum PageRanks were able to have faster stabilization and consistent PageRank ordering by adding a little noise during the computation of the Kossakowski-Lindblad master equation.

M. Bidlo and P. Zufan [10] performed a comparative study on the evolutionary design of quantum operators. Genetic Algorithm and Evolution Strategy are applied, each in four different setups, and evaluated on three case studies: the 2-qubit Controlled-NOT gate, 3-qubit entanglement operator and 4-qubit detector of an element with the maximum amplitude. The newly applied QR decomposition achieved 100% success in 3-qubit entanglement using both GA and ES and the best statistical evaluation in case of the 4-qubit operator. Quantum computing has superior computational strengths than the classical computer and NP-hard problems have been tackled with this method.

Graph partitioning (NP-Hard) graph problem has been solved by U. Chukwu *et al.* [11] utilizing two quantum-ready methods of QUBO (quadratic unconstrained binary optimization) and constrained-optimization sampler. Both approaches often delivered better partition than the purpose-built classical graph partitioners.

Stuart M. Harwood *et al.* [12] focus on variational quantum eigensolver (VQE) which is a hybrid quantum-classical algorithm. The authors adopted variational adiabatic quantum computing (VAQC) to propose an improved VQE method that continuously parameterized Hamiltonian via the quantum circuit. The proposed technique VAQC has the ability to successfully find good initial circuit parameters to initialize VQE. The method was evaluated with two examples from quantum chemistry combined with other techniques that provide more accurate solutions than conventional VQE, for the same amount of effort.

S. Boyapati, S. R. Swarna and A. Kumar [13] evaluated the performance of a quantum computed prediction mode using Quantum Neural Networks (QNN). This computational mechanism using quantum computing and the neural network will track the live operations and form the dynamic route changes in the real-time scenario. This real-time scenario worked with a 95% accuracy rate with its accuracy differing based on the number of connecting nodes being considered.

Rosa M. Gil Iranzo et al. [14] addresses the limitations of quantum computing interfaces that facilitate learning the emerging paradigm. The authors proposed a layer to create proper learning environments for performing calculations without facilitating the understanding of the principle of quantum computing concepts. The proposed work focuses on Human-centered computing that shall facilitate various levels including high school, university, and the research level. This research is novel around the domain of design of quantum computing interfaces integrating science and technology.

## III. RESEARCH METHODOLOGY

The systematic literature review [15], [16] has been conducted to find the current innovations that are either completely new or modification of existing approaches for the study on the Evolution of Quantum Computing, depicted in Figure 1. A "Search Process" was implemented to acquire research papers that address our topic of study. Thus, specific search strings were applied during our analysis in scientific databases which contained the keywords, "Quantum Computing", "Quantum Computing Evolution" and "Quantum Computing Tools".

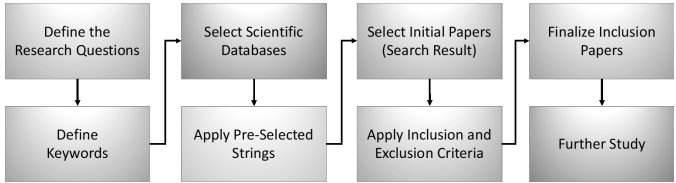

Figure 1: Systematic Literature Mapping Process [17]

The scientific databases that were used for procuring these papers including: (i) IEEE Xplore, (ii) ScienceDirect, (iii) ACM, (iv) Springer Link, and (v) ResearchGate. We adopt a screening process to find the most relevant papers by studying the paper title followed by reading and understanding the abstract and conclusion from screened papers. An exclusion and inclusion process based on (i) duplicate papers (ii) full-text availability, and (iii) papers that are not related to Quantum Computing was conducted to prune off research papers that had aspects that were not related to our literature review as well as duplicates that appeared during the initial search. Table III. displays the details of the inclusion and exclusion process.

TABLE II: GENERALIZED TABLE FOR SEARCH CRITERIA

| Scientific Database | Initial Keyword Search | Total Inclusion |
|---|---|---|
| *IEEE Xplore* | 109 | 4 |
| *ScienceDirect* | 50 | 2 |
| *ACM* | 50 | 2 |
| *Springer Link* | 25 | 0 |
| *Research Gate* | 17 | 0 |
| **Total** | **251** | **8** |

TABLE IIII: OVERVIEW OF EXCLUSION AND INCLUSION

| Category | Condition (Exclusion) | Condition (Inclusion) |
|---|---|---|
| *Title* | Does not include inclusion topics | Quantum Computing Evolution, Tools, Methods |
| *Duplicate Papers* | Similar papers in multiple scientific databases | Papers are not duplicated in different scientific databases |
| *Relativity* | Studies that do not cover expected domain | Proposed approaches reflect Quantum Computing |
| *Text Availability* | Papers are not available fully and not in English | Papers that are available in the full format & in English |

Firstly, the filtration procedure had a time constraint that allowed research papers published from the years 2016 to 2022. Furthermore, additional filters were placed in each database to narrow our search of relevant research materials. IEEE Xplore included Conferences and Journals while ScienceDirect required us to select Computer Science as the subject area and research articles for article type. Springer Link and Research Gate did not provide any unique or relevant topics of research. We also included ACM scientific database where we included 2 papers for this study. Total 251 research papers were found during the initial search but an in-depth screening process that accounted for the publication title, abstract, experimental results and conclusions shortened the list to 8 papers for our study.

## IV. PLATFORM USED FOR QUANTUM COMPUTING

Quantum Platform (QP) or Quantum Computer Platform (QCP) is a family of lightweight, open-source software frameworks for building responsive and modular real-time embedded applications in Quantum Computing [18]. Quantum Computer Platform consists of two layers: Quantum Computing Layer and Classical Computing Layer [19], [20] depicts in Figure 2. In this section, we present our findings on platform used for quantum computing with a research question, what are the platforms used for quantum computing in the literature?

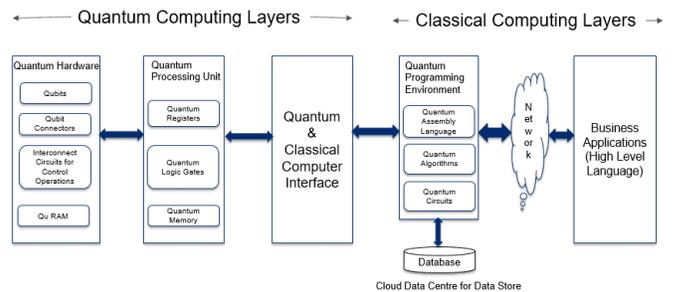

Figure 2: Quantum Computer Platform Architecture

### A. The Quantum Computing Layers

An optimal set of hyperparameters allows performance improvement as well as avoid performance issues like overfitting. The Quantum hardware covers Qubits which are surrounded by superconducting loops for the physical realization of Qubits. It also consists of the internally connected circuitry for Qubit control operations. Quantum Processing Unit includes Quantum registers, logic gates, and memory. The

Quantum-Classical Interface houses the required hardware and software in order to provide interfacing between the classical computers and a Quantum Processing Unit (QPU). Lastly, the Classical Computing Layer includes the final components- Quantum Programming environment, Cloud data Centre and Business Applications.

### B. Characteristics of Quantum Computer Platform

- Low-level Programming: The Quantum Computers currently in use are built on low-level programming. They are based on quantum logical gates and handle computational steps to execute in QPU.

- Heterogeneous: In QCP the technical specifications are heterogeneous in nature for both software and hardware. Some examples of QCP (or QP) are IBM, Microsoft, D-wave, Google.

- Remote software development and deployment: All the QCP vendors provide Quantum Computing software development frameworks for leveraging quantum processors that can only be accessed remotely from the cloud. Only a small part of the programming tool stack is deployed on the local machines. So the programmers access quantum software remotely for development and testing.

- Quantum algorithms: The popular algorithms help in gaining speed and communicating with other computing tasks that are running on QCP. Additionally, programmers need to either identify or design suitable algorithms to solve the problems in hand.

- Portability of Software: The software developed by the QCP owners are currently native in nature. This software always follow its own standards, proprietary programming API and predefined tools. Examples of software will be found in the next segments of this paper.

### V. CIRCUIT SIMULATOR FOR QUANTUM COMPUTING

The working procedure of a quantum circuit is shown with the following diagrams of a simulator called "Quirk" [21], one of the most used simulators for Quantum Computing. In the circuit of Figure 3, there are two |0⟩|0⟩ qubits. When the gates are dragged onto these circuits the output changes accordingly [22].

The Hadamard gate or H-gate [23] is a quantum logic gate. It redistributes the probability of all the input lines. As a result, the output lines have an equal chance of being 0 and 1. Dragging an H-gate onto one of the |0⟩|0⟩ circuits from Figure 3 will give the circuit in Figure 4. Implementing the H-gate in this circuit the output has a 50% chance of being measured ON, or 1.

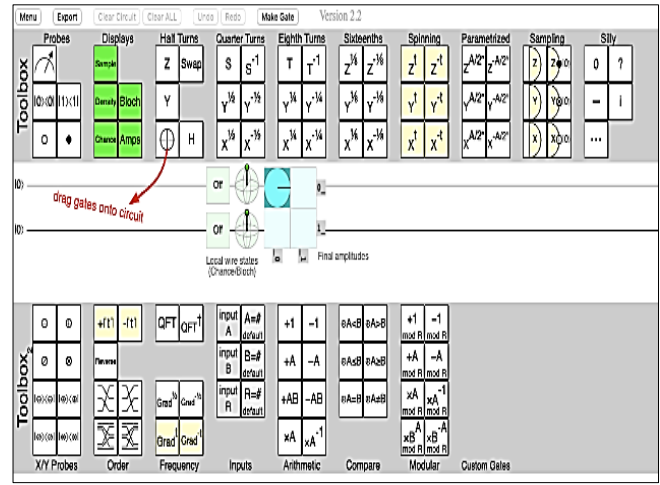

Figure 3: Quantum Circuit with Quirk

Adding a new gate to the circuit of Figure 3 will give the circuit in Fig. 4. The new gate is called the Pauli X gate, classical the quantum equivalent of the NOT gate. This gate flips the input state, so a 0 as input becomes 1 as output, and vice versa. From the circuit of Figure 5 that is visible that the chance of measuring a 1 is 100%. Some common quantum logic gates with their associated matrices are shown in Table IV below.

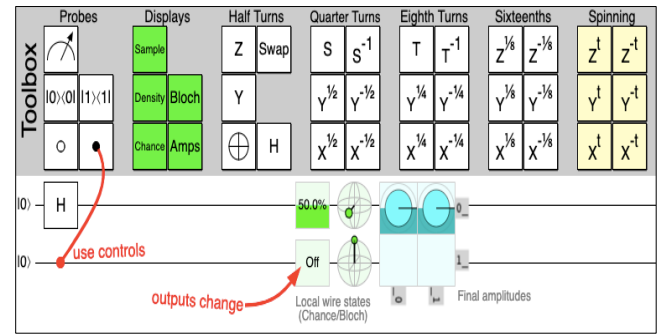

Figure 4: Use of Hadamard Gate on the circuit

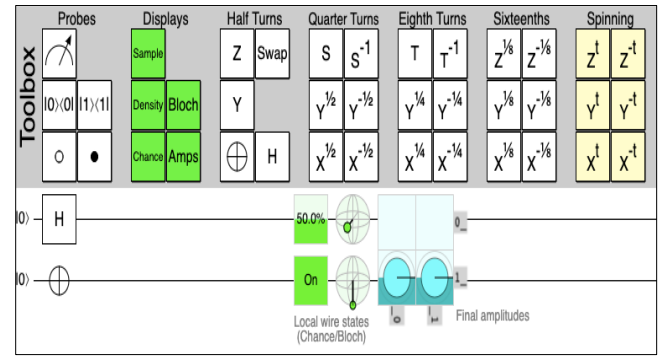

Figure 5: Use of Pauli's X-gate on the circuit

TABLE IV: COMMON QUANTUM LOGIC GATES WITH THEIR ASSOCIATED MATRICES

| Gate | Matrix |
|---|---|
| Pauli-X | $\begin{vmatrix} 0 & 1 \\ 1 & 0 \end{vmatrix}$ |
| Pauli-Y | $\begin{vmatrix} 0 & -i \\ i & 0 \end{vmatrix}$ |
| Pauli-Z | $\begin{vmatrix} 1 & 0 \\ 0 & -1 \end{vmatrix}$ |
| Hadamard (H) | $\frac{1}{\sqrt{2}} \begin{vmatrix} 1 & 1 \\ 1 & -1 \end{vmatrix}$ |
| Phase (S,P) | $\begin{vmatrix} 1 & 0 \\ 0 & i \end{vmatrix}$ |
| π/8 (T) | $\begin{vmatrix} 1 & 0 \\ 0 & e^{i\pi/4} \end{vmatrix}$ |
| Controlled NOT (CNOT, CX) | $\begin{vmatrix} 1 & 0 & 0 & 0 \\ 0 & 1 & 0 & 0 \\ 0 & 0 & 0 & 1 \\ 0 & 0 & 1 & 0 \end{vmatrix}$ |
| Controlled Z (CZ) | $\begin{vmatrix} 1 & 0 & 0 & 0 \\ 0 & 1 & 0 & 0 \\ 0 & 0 & 1 & 0 \\ 0 & 0 & 0 & -1 \end{vmatrix}$ |
| SWAP | $\begin{vmatrix} 1 & 0 & 0 & 0 \\ 0 & 0 & 1 & 0 \\ 0 & 1 & 0 & 0 \\ 0 & 0 & 0 & 1 \end{vmatrix}$ |
| Toffoli(CCNOT, CCX, TOFF) | $\begin{vmatrix} 1 & 0 & 0 & 0 & 0 & 0 & 0 & 0 \\ 0 & 1 & 0 & 0 & 0 & 0 & 0 & 0 \\ 0 & 0 & 1 & 0 & 0 & 0 & 0 & 0 \\ 0 & 0 & 0 & 1 & 0 & 0 & 0 & 0 \\ 0 & 0 & 0 & 0 & 1 & 0 & 0 & 0 \\ 0 & 0 & 0 & 0 & 0 & 1 & 0 & 0 \\ 0 & 0 & 0 & 0 & 0 & 0 & 0 & 1 \\ 0 & 0 & 0 & 0 & 0 & 0 & 1 & 0 \end{vmatrix}$ |

## VI. TOOLS AND SOFTWARE FOR QUANTUM COMPUTING

Some of the tools and software used for Quantum Computing are discussed in this section with a research question, what are the tools and software are being used for quantum computing in the literature?

### A. Cirq

Cirq [24], [25] is an open-source Python library for writing, manipulating and optimizing Noisy Intermediate Scale Quantum (NISQ) circuits, and also for running them against quantum computers and simulators illustrate in Figure 6. Moreover, it can be used with OpenFermion-Cirq which is a platform for developing quantum algorithms for chemistry problems. Cirq is not an official Google product, but Google AI Quantum Team is promoting it.

### B. TensorFlow Quantum

TensorFlow Quantum (TFQ) [26] is a quantum machine learning library that is being used for prototyping of hybrid quantum-classical machine learning models by Google illustrate in Figure 7. It works with Cirq [27] to provide quantum computing primitives compatible with existing TensorFlow APIs, along with high-performance quantum circuit simulators.

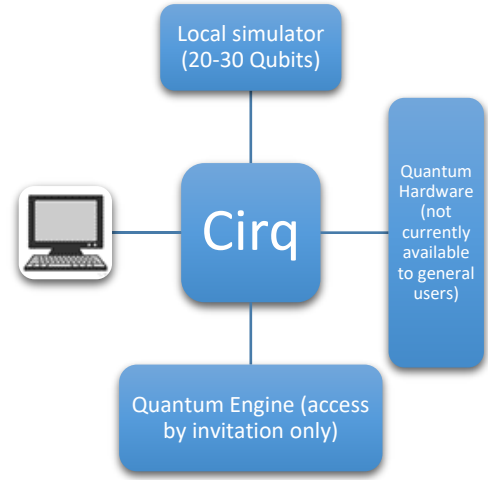

Figure 6: A general overview of Cirq

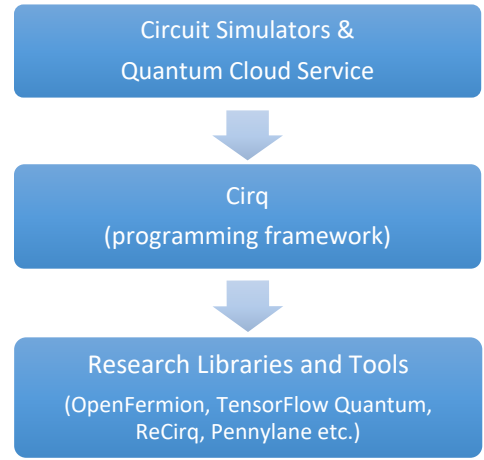

Figure 7: How TensorFlow Quantum works with Cirq

### C. ProjectQ

ProjectQ [28] is an open-source software framework that allows users to implement quantum programs in Python using a powerful and intuitive syntax (Fig 8). After that it can translate these programs to any type of back-end, either a simulator running on any classical computer or an actual quantum chip including the IBM Quantum Experience platform.

### D. CirqProjectQ

CirqProjectQ [29] is a port between ProjectQ and Cirq that provides two main functions. As the first function, it has a

ProjectQ backend that converts a ProjectQ algorithm to a cirq circuit. Secondly, it can decompose ProjectQ common gates to native Xmon gates to simulate a Google quantum computer with ProjectQ.

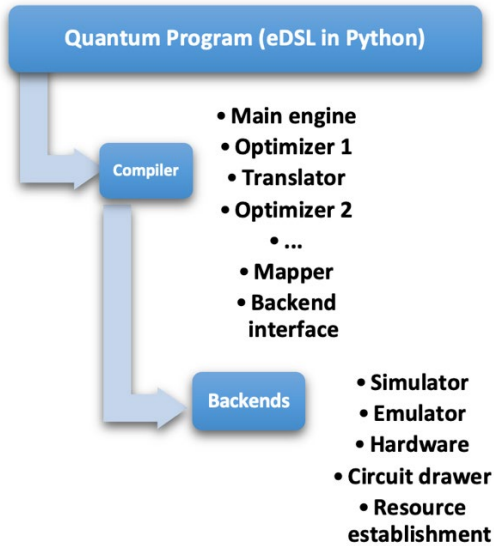

Figure 8: Working procedure of ProjectQ

### E. Microsoft Quantum Development Kit

Microsoft Quantum Development Kit [30], [31] appears to supercede their earlier LIQUi|> software. This kit features a new quantum programming language Q#. It works with integrating the Visual Studio development environment (Figure 9 and Figure 10).

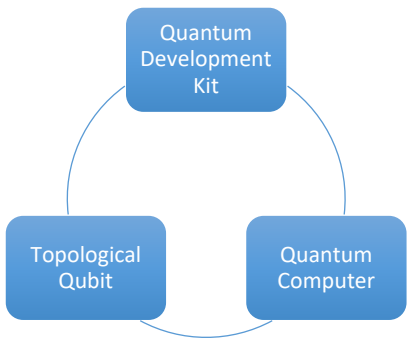

Figure 9: Working procedure of Microsoft Quantum Development Kit

### F. IBM Quantum Experience

IBM's 5 qubit gate-level quantum processor on the web allows the users to apply to get access to it. The IBM Quantum Experience [32], [33] website shows four modules, a short tutorial with instructions to use it, a quantum composer to configure quantum gates for the qubits, a simulator to simulate the configuration before running it on the actual machine, and finally access to the machine itself to run the configuration and view the results (Figure 11). It has an associated software API called QISKIT.

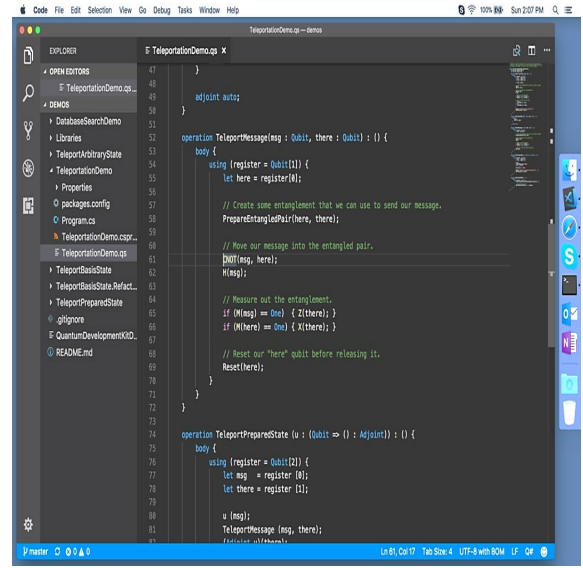

Figure 10: IDE of Microsoft Quantum Development Kit (integrated with Microsoft Visual Studio)

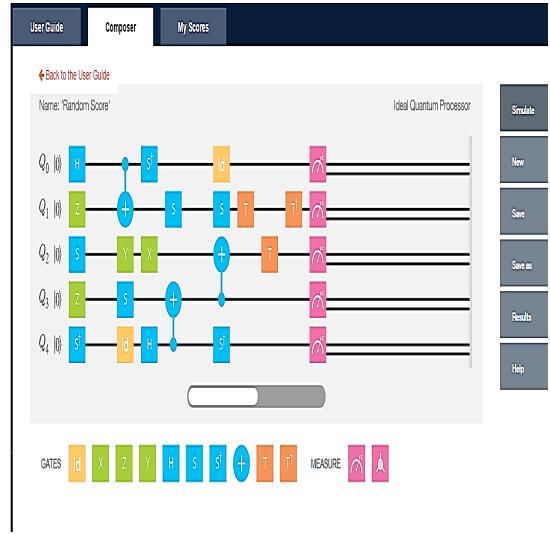

Figure 11: How IBM Quantum Experience works

### G. Rigetti Forest and Cloud Computing Services (QCS)

The Rigetti Forest suite [34], [35] consists of a quantum instruction language Quil, an open source Python library pyQuil, a library of quantum programs called Grove and a simulation environment called QVM (Quantum Virtual Machine). QCS provides a virtual classical computing environment alongside the Rigetti quantum hardware. It comes pre-configured with Rigetti's Forest SDK and provides the users with a single access point to the QVM and QPU backends.

### H. CAS-Alibaba Quantum Computing Laboratory

The CAS-Alibaba Quantum Computing Laboratory [36] has built several superconducting quantum computers. Their hardware systems are available through an online interface for the users to write quantum circuits, execute them, and download the results over the cloud using a GUI.

### I. Quantum Computing Playground

The Quantum Computing Playground [37] is a Chrome Experiment or web app (Figure 12) uses WebGL to simulate up to 22 qubits on a GPU. Inside it the users get a basic IDE to write, compile and run the code; along with some example algorithms (Grover's, Shor's). Also, a debugger and 3D quantum state visualization tool are there, so users can see what's going on inside the little quantum computer. QScript is the programming language used here, and it is similar to Bash-like scripting languages.

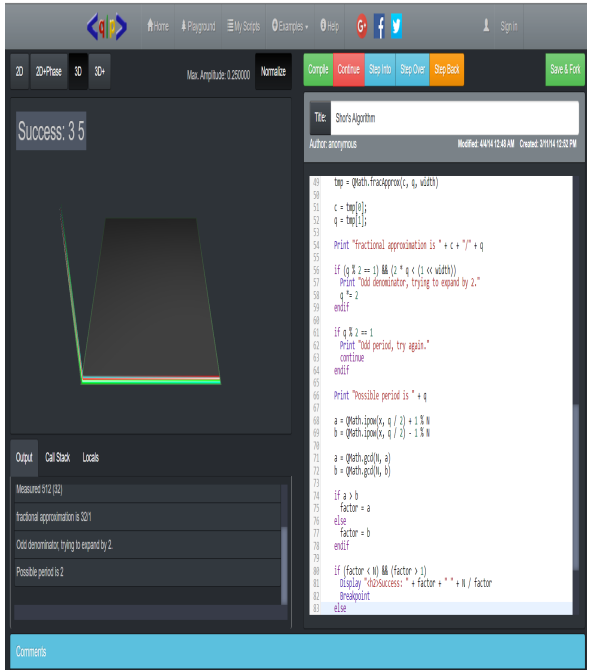

Figure 12: How Quantum Computing Playground works (simulating the example of Shor's algorithm)

### J. Strawberry Fields

Strawberry Fields [38] is an open-source quantum programming architecture for quantum machine learning depicts in Figure 13.

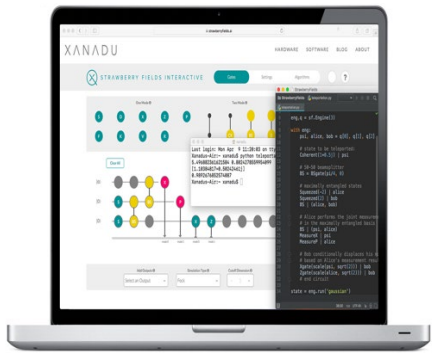

Figure 13: Display How Strawberry Fields works

Building with Python language and consists of a full-stack library for design, simulation, optimization and quantum machine learning of several paradigmatic algorithms, such as-teleportation, (Gaussian) boson sampling, instantaneous quantum polynomial, Hamiltonian simulation and variational quantum circuit optimization.

## VII. CURRENT SITUATION OF THE FIELD AND ITS CHALLENGES

Today's Quantum Computers take up an entire room, but their capabilities are all really small-scale till now. They possess less than 100 Qubits each which does not seem enough for the tasks they are up to. Currently, the Quantum Computer with the highest number of Qubits is China's Zuchongzhi with 66 Qubits [39], [40]. It is able to perform a sampling task in 1.2 hours that would take eight years for a Classical Computer to complete. At present, about 46 countries are engaged in national or international Quantum research works and developments. Most of these actions are found to happen in academia and industry [41], [42].

Lack of good software leads to technological challenges in Quantum Computation including limited qubit connectivity, too low gate fidelities, and the requirement of large amounts of qubits for error correction. Lack of collaboration and exchange between industry and academia is also a major issue in the advancement of Quantum Computing. Furthermore, such computers operate at temperatures close to zero, and maintaining such a low temperature is always a big challenge. Today, computers with 70 Qubits fall short of the requirement of one million Qubits to make economically feasible and viable Quantum Computers.

## VIII. RECOMMENDATIONS FOR FUTURE RESEARCH

Cloud-oriented Quantum Computing has the potential to overtake future business initiatives and technologies including cryptography, machine learning (ML), and artificial intelligence (AI) [43]–[46]. It seems plausible because of the almost unlimited memory spaces available in clouds. Besides, shared hardware could be proved helpful in solving complex tasks with Quantum algorithms but by using a Classical Computer as its base.

Looking at the possibilities, Quantum AI tools may provide the world with autonomous weapons and mobile platforms [47]–[49]. For example, drones made with Quantum AI tools can achieve heightened sensing, navigation, and positioning options in GPS-denied areas, as well as altering the course of operation to avoid enemy countermeasures.

In addition, a new internet possibility with Quantum devices emerged called the Quantum internet, which is separate from the internet and links Quantum devices together using entanglement. It significantly increases the connectivity, security, and speed of the internet and shows the potential of the super-secure communication infrastructure that protects Quantum-internet connected devices from cyberattacks to serve in the field of cryptography. For instance, some scientists in the Netherlands entangled three one-qubit devices in this manner, and they successfully communicated and stored information in a theoretically unhackable manner.

Moreover, Quantum cryptography, ML and AI tools can be combined to improve intelligence service and its analysis [50], [51]. Such intelligence services are supposed to be able to break 2048-bit RSA encryption in 8 hours or even less- a task that would require the world's fastest supercomputers around 300 trillion years to complete with brute-force methods. Quantum computers might demand almost 20-million Qubits to perform it. Advances in this field show possibility of such machines in 25 years. There is a chance of adversarial use of such computing which can risk national and international security if advances in Quantum decryption outrun advances in Quantum encryption. With the advancement of Quantum Computing, ML and AI problems could be solved in a practical amount of time- reduced from hundreds of thousands of years to a few seconds.

## IX. Conclusion

Quantum Computing harnesses the phenomena of quantum mechanics to solve complex problems that today's most powerful supercomputers cannot solve. In this paper, we reviewed the evolution of Quantum Computing where tremendous efforts and progress mark a significant milestone in solving real-world problems in recent years. We also present the progress of various frameworks, tools, software, and platforms that facilitate quantum computing. Finally, we discussed the current essence, challenges, and provide a research scope for future research.

## Acknowledgement

The work is partially supported by the U.S. National Science Foundation Awards #2100115, #1723578, #1723586. Any opinions, findings, and conclusions or recommendations expressed in this material are those of the authors and do not necessarily reflect the views of the National Science Foundation.